# Self-supervised iRegNet for the Registration of Longitudinal Brain MRI of Diffuse Glioma Patients

Ramy A. Zeineldin [1,2,3], Mohamed E. Karar[2], Franziska Mathis-Ullrich[3], Oliver Burgert[1]

[1] Research Group Computer Assisted Medicine (CaMed), Reutlingen University, Germany
[2] Faculty of Electronic Engineering (FEE), Menoufia University, Egypt
[3] Health Robotics and Automation (HERA), Karlsruhe Institute of Technology, Germany
`Ramy.Zeineldin@Reutlingen-University.DE`

**Abstract.** Reliable and accurate registration of patient-specific brain magnetic resonance imaging (MRI) scans containing pathologies is challenging due to tissue appearance changes. This paper describes our contribution to the Registration of the longitudinal brain MRI task of the Brain Tumor Sequence Registration Challenge 2022 (BraTS-Reg 2022). We developed an enhanced unsupervised learning-based method that extends the iRegNet. In particular, incorporating an unsupervised learning-based paradigm as well as several minor modifications to the network pipeline, allows the enhanced iRegNet method to achieve respectable results. Experimental findings show that the enhanced self-supervised model is able to improve the initial mean median registration absolute error (MAE) from 8.20 ± 7.62 mm to the lowest value of 3.51 ± 3.50 for the training set while achieving an MAE of 2.93 ± 1.63 mm for the validation set. Additional qualitative validation of this study was conducted through overlaying pre-post MRI pairs before and after the deformable registration. The proposed method scored 5th place during the testing phase of the MICCAI BraTS-Reg 2022 challenge. The docker image to reproduce our BraTS-Reg submission results will be publicly available.

## 1 Introduction

Glioblastoma (GBM), and diffuse glioma, are the most common and aggressive malignant primary tumors with high and heterogeneous infiltration rates [1]. The registration of longitudinal brain Magnetic Resonance Imaging (MRI) scans is crucial in the treatment and follow-up procedures of brain tumors to find map correspondences between pre-operative and post-recurrence. This would support research into the early detection of tumor infiltration and subsequent tumor [2]. Therefore, an automatic, fast, robust fusion of follow-up with the pre-operative MRI scans becomes highly important to assist in the early detection of tumor recurrence. However, the registration of MRI brain glioma patients is still a complex and challenging problem due to the inconsistent intensity and missing correspondences between both scans, especially with large deformations caused by large tumors (Glioma grade III and IV).

Over the past years, many approaches have been applied to medical image registration that can be classified into classical and learning-based approaches [3, 4]. Classical or non-learning methods are formulated as an iterative pair-wise optimization problem that requires proper feature extraction, choosing a similarity measurement, defining the used transformation model, and finally an optimization mechanism to investigate the search space. Over time, an extensive literature has developed using diverse combinations of the aforementioned elements [5-9]. Still, the traditional iterative process is computationally expensive, requiring long processing times ranging from tens of minutes to hours even with an efficient implementation on a regular central processing unit (CPU) or modern graphical processing unit (GPU).

To overcome the limitations of classical methods, learning-based approaches have been proposed in recent years. Learning methods formulate the classical optimization problem into a problem of cost function estimation. Instead of finding the map correspondence for every input MRI scanning separately, learning approaches make a general optimization over all the training datasets [4]. Recently, deep learning has been widely adopted in various medical image analysis tasks outperforming other methods [10]. Supervised deep learning methods were initially proposed [11-13] to learn similar features from the training data using different imaging modalities. Then, unsupervised learning was developed as a demand for faster registration procedures and to eliminate the challenges related to ground truth data generation and optimization techniques [14-17]. In general, once the deep learning networks are trained, they can provide a faster registration than classical optimization methods, without the need for fine-tuning parameters at the test time, in addition to being more robust to outliers.

In this paper, we propose a fully automatic, patient-specific registration approach for pre- and post-operative brain MRI sequences of only a single modality using iRegNet [18]. In particular, we introduce an unsupervised approach of iRegNet (see Fig. 1) in which only moving and fixed MRI pairs are utilized. Then, our proposed method optimizes deformation fields directly from input images using backpropagation. Extensive experiments of our model on the BraTS-Reg challenge data of 160 patients show that the proposed method can provide accurate results with the advantage of having a very fast runtime.

The remainder of the paper is organized as follows: Section 2 describes the BraTS-Reg 2022 dataset and our patient-specific registration framework. Qualitative and quantitative evaluations are presented in Section 3; Section 4 concludes the paper with an outlook on future work.

## 2 Material and Methods

### 2.1 Dataset

The BraTS-Reg 2022 dataset [19] comprises 250 patient-specific pairs of pre-operative and follow-up brain multi-institutional MRI scans. For each patient, i) native T1-weighted (T1), ii) contrast-enhanced T1 (T1ce), iii) T2-weighted (T2), and iv) T2 Fluid

Attenuated Inversion Recovery (FLAIR) sequences are provided for the pre-operative and follow-up with a time-window in the range of 27 days-37 months. Reference landmark annotations for the validation set are not made available to the participants. Instead, participants can use the online evaluation platform[1] to evaluate their models and compare their results with other teams on the online leaderboard[2].

Standard pre-processing techniques were applied such as rigid registration to the same anatomical template, resampling to the same isotropic resolution ($1mm^3$), skull removal, and brain extraction. Following these pre-processing steps, we applied the image cropping stage where all brain pixels were cropped. Afterward, z-score normalization was applied by subtracting the mean value and dividing it by the standard deviation individually for each input MRI image.

## 2.2 Medical Image Registration

Medical image registration is the process of aligning two or more sets of imaging data acquired using mono- or multi-modalities into a common coordinate system. Let $\boldsymbol{I_F}$ and $\boldsymbol{I_M}$ denote the fixed and the moving images, respectively, and let $\boldsymbol{\phi}$ be the deformation field that relates the two images. Then, our goal is to find the minimum energy function $\boldsymbol{\varepsilon}$ as:

$$\boldsymbol{\varepsilon = S(I_F, I_M \cdot \phi) + R(\phi)} \tag{1}$$

where $(\boldsymbol{I_M \cdot \phi})$ is the moving image $\boldsymbol{I_M}$ warped by the deformation field $\boldsymbol{\phi}$, the dissimilarity metric is denoted by $\boldsymbol{S}$, and $\boldsymbol{R(\phi)}$ represents the regularization parameter. In this work, follow-up and pre-operative MRI scans are utilized as the moving and fixed images, respectively, since our goal is to reflect the brain shift in the follow-up MRI data.

## 2.3 iRegNet workflow

Figure 1 presents an outline of the baseline iRegNet registration method. iRegNet consists of two steps: First, $\boldsymbol{I_F}$ and $\boldsymbol{I_M}$ are fed into our convolutional neural network (CNN) that then predicts $\boldsymbol{\phi}$. Second, $\boldsymbol{I_M}$ is transformed into a warped image $(\boldsymbol{I_M \cdot \phi})$ using a spatial re-sampler. Further details are described as follows.

**CNN Architecture.** The developed CNN architecture utilized in experiments is based on U-Net [20, 21]. Using backpropagation, which is a feedback loop that estimates the network weighting parameters, the network can automatically learn the optimal features and the deformation field. The network contains two main paths: a feature extractor (or encoder) as well as a deformation field estimator (or decoder). 3D convolutions are applied in both encoder and decoder parts instead of the 2D convolutions used in the original U-Net architecture. The encoder consists of two consecutive 3D convolutional layers, each followed by a rectified linear unit (ReLU) and 3D spatial max pooling. A stride of 2 is employed to reduce the spatial dimension in each layer by half, similar to the traditional pyramid registration scheme. In the decoding path, each step consists of a 3D up-sampling, a concatenation with the corresponding features from the encoder, 3D up-convolutions, and a batch normalization layer, followed by a rectified linear unit (ReLU). Finally, a 1 x 1 x 1 convolution layer is applied to map the resultant feature vector map into $\boldsymbol{\phi}$.

---

[1]  https://ipp.cbica.upenn.edu/
[2]  https://www.cbica.upenn.edu/BraTSReg2022/lboardValidation.html/

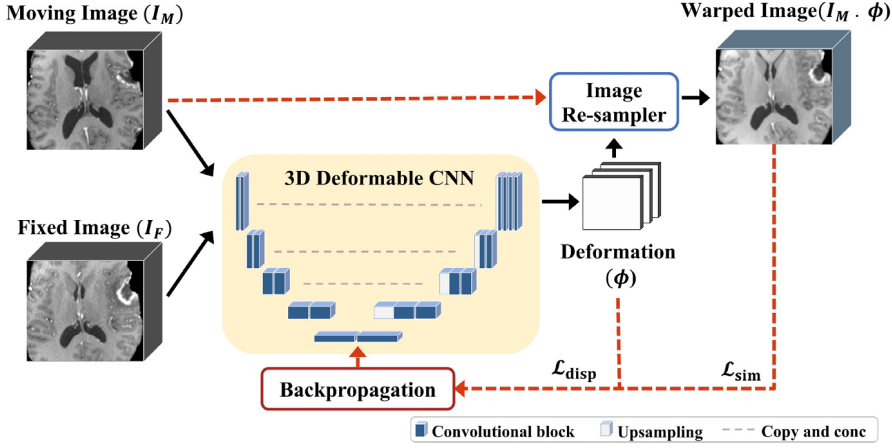

**Fig. 1.** An overview of the iRegNet workflow for 3D post- to pre-operative MRI image deformable registration. Dashed *red* arrows show the processes applied in the training stage only [18].

**Self-supervised Learning.** In contrast to the original iRegNet where supervised learning was applied, we incorporated self-supervised learning to compute the optimal deformation field $\widehat{\boldsymbol{\phi}}$ corresponding to the smoothness regularization. This model uses only the input MRI volume pair, and the registration field is computed accordingly by the CNN network. Formally, this task is defined as:

$$\widehat{\boldsymbol{\phi}} = \arg\min_{\boldsymbol{\phi}} \mathcal{L}_{sim}(\mathbf{I_F}, \boldsymbol{\phi}.\mathbf{I_M}) + R(\boldsymbol{\phi}) \qquad (2)$$

where $\mathcal{L}_{sim}$ computes the image similarity between the warped image $(\boldsymbol{\phi} \, . \, \boldsymbol{I_M})$ and the fixed image $\boldsymbol{I_F}$,

**Loss Function.** Owing to the applied two-step approach, the overall loss function $\mathcal{L}_{overall}$ has two components, as shown in Equation (3). $\mathcal{L}_{disp}$ corresponds to the deformation field gradient error.

$$\mathcal{L}_{overall} = \mathcal{L}_{sim} + \mathcal{L}_{disp} \qquad (3)$$

where $\mathcal{L}_{sim}$ employs the similarity metric of the local normalized correlation coefficient (*NCC*), which is calculated as follows:

$$\mathcal{L}_{sim} = \mathbf{NCC}(\mathbf{I_F}, \boldsymbol{\varphi}. \mathbf{I_M}) = \frac{1}{N} \sum_{p \in X} \frac{\sum_i (\mathbf{I_F}(p) - \overline{\mathbf{I_F}(p)}) \sum_i (\boldsymbol{\varphi}. \mathbf{I_M}(p) - \overline{\boldsymbol{\varphi}. \mathbf{I_M}(p)})}{\sqrt{\sum_i (\mathbf{I_F}(p) - \overline{\mathbf{I_F}(p)})^2} \sqrt{\sum_i (\boldsymbol{\varphi}. \mathbf{I_M}(p) - \overline{\boldsymbol{\varphi}. \mathbf{I_M}(p)})^2}} \quad (4)$$

where $(\boldsymbol{\varphi}. \boldsymbol{I_M(p)})$ and $\boldsymbol{I_F(p)}$ are the voxel intensities of a corresponding patch $\boldsymbol{p}$ in the warped image and the fixed truth, respectively, whereas $\overline{(\boldsymbol{\varphi}. \boldsymbol{I_M(p)})}$ and $\overline{\boldsymbol{I_F(p)}}$ are the mean pixel intensities for both images. $\mathcal{L}_{disp}$ measures spatial gradients differences in the predicted displacement $\boldsymbol{d}$ as follows:

$$\mathcal{L}_{disp} = \sum_{p \in X} \|\nabla \boldsymbol{d(p)}\| \quad (5)$$

## 3 Experimental Results

### 3.1 Experimental Setup

To ensure computational efficiency for the GPU, MRI scans for each patient were center cropped to $160 \times 192 \times 160$ pixels. For training and validation, sets, respectively, the BraTS-Reg dataset was randomly split into 112 (80%) and 28 (20%) volumes. Another 20 MRI volumes were provided by the BraTS-Reg organizers as online validation set with landmarks provided only for the fixed MRI scans. Finally, we performed an affine alignment on moving and fixed MRI volumes using the BRAINSFit toolkit [9] to focus on the non-linear misalignment between volumes. For the experiments, our model was implemented in Python 3.7 using the TensorFlow 2.4 library. The experiments were run on an AMD Ryzen 2920X (32M Cache, 3.50 GHz) CPU with 64 GB RAM and a single NVIDIA GPU (RTX 3060 12 GB or RTX 2080 Ti 11 GB). The ADAM optimizer[22] with an initial learning rate of 1e$^{-4}$ and a batch size of 2 was used.

To compare with other studies, the mean target registration error (mTRE), which represents the average distance between the corresponding landmarks in each pre-post MRI pair before and after registration, was used. In addition, the proposed method was evaluated by the online submission platform using the following metrics, namely Median Absolute Error (MAE), Robustness, and smoothness of the displacement field.

### 3.2 Ablation Study

To explore the MRI modality which achieves the best performance for the task of longitudinal registration, an ablation study has been carried out. The BRAINSFit toolkit was utilized to perform affine alignments on moving and fixed MRI volumes. As listed in Table 1, T1ce has obtained the overall best results on the validation dataset in terms of the mean and median MAE scores while FLAIR achieved the best robustness. Therefore, in our experiments, we only use the T1ce volumes from each patient. Theoretically, using multiple modalities could increase the accuracy of the image registration, and this would be further investigated in future work.

**Table 1.** The ablation study of MRI modalities on the BraTS-Reg 2022 validation cases. Bold highlights the best scores.

| Modality | $MAE_{median}$ | $MAE_{mean}$ | Robustness |
|----------|--------------|------------|------------|
| Initial | 8.20 | 8.65 | - |
| T1 | 4.74 | 5.65 | 0.66 |
| T1ce | **4.35** | **5.23** | 0.62 |
| T2 | 4.85 | 5.60 | 0.63 |
| FLAIR | 4.64 | 5.40 | **0.67** |

### 3.3 Registration Results

Figure 2 shows example results from three patients, where the registration of post- to pre-operative MRI scans is achieved using the self-supervised iRegNet method and the comparing baseline method. From the visual results, it can be seen that warped MRI scans are significantly improved after applying iRegNet. Note that, Fig. 2 (c) shows the FLAIR scans for the follow-up MRI images, only for visualization purposes, to better depict the surgically imposed cavities of these illustrated examples. All the applied registration methods use only the T1ce modality as discussed in Section 3.2.

Moreover, Table 2 reports the registration performance of the proposed method as well as the baseline on the BraTS-Reg challenge validation database. The baseline denotes the BRAINSFIT affine transformation between the full-resolution images of pre-operative and follow-up MRI. Compared with the affine method, our proposed self-supervised method effectively improves registration performance. It is notable that the average runtime of the proposed method is 1 second and does not require any manual interaction or supervision. Besides, only one sequence (T1ce) is required in our case.

The statistics of the paired landmark errors before and after the registration are displayed in Fig. 3. For the training database, our model reduced the initial mean MAE (computed by the evaluation platform) from $8.20 \pm 7.62$ mm to $3.51 \pm 3.50$ mm. Similarly, an MAE of $2.93 \pm 1.63$ mm was achieved on the validation database which has an initial $7.80 \pm 5.62$ mm. This result highlights that our method delivers significantly better results than both initial alignment and affine registration.

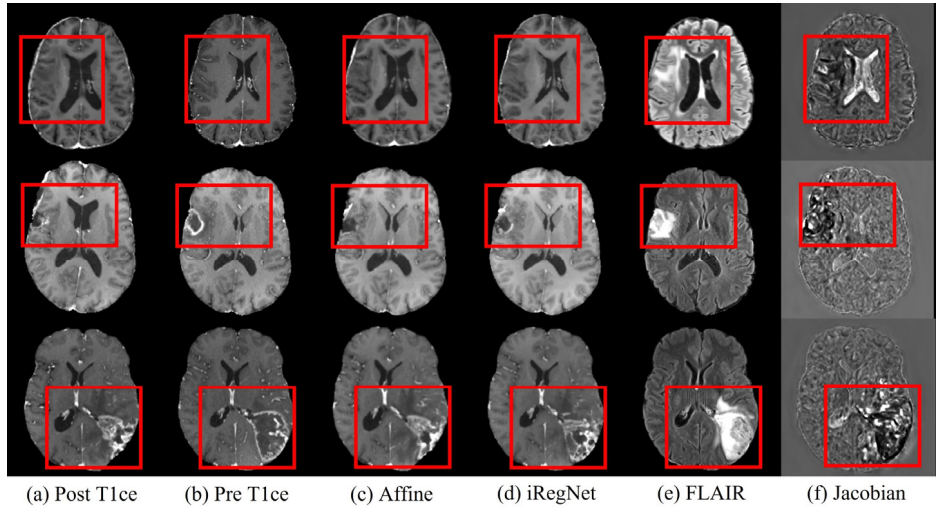

| (a) Post T1ce | (b) Pre T1ce | (c) Affine | (d) iRegNet | (e) FLAIR | (f) Jacobian |

**Fig. 2.** Example registration results from three validation cases (patients 141, 148, and 152). From left to right: (a) and (b) the post- and pre-operative MRI T1ce, (c) the follow-up to pre-operative affine registration of BRAINSFit, (d) the follow-up to pre-operative deformable registration of our iRegNet, (e) the pre-operative FLAIR scans, only for visualization purposes, and (f) determinant of the Jacobian of the displacement field are shown, respectively. The red box highlights regions of major differences.

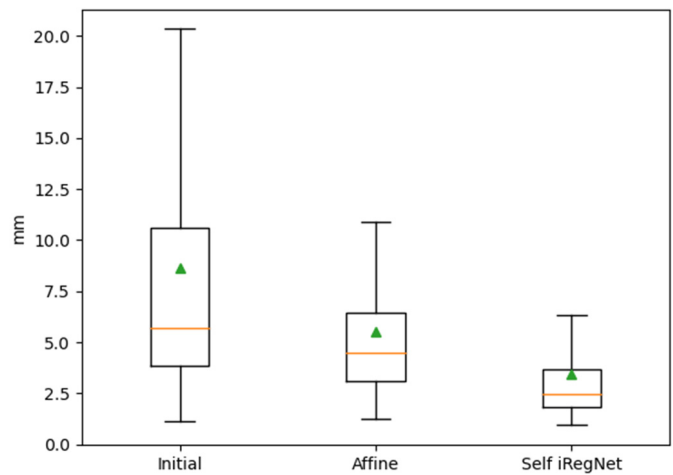

**Fig. 3.** Boxplots of the mean landmark errors. For each method, the landmark errors are computed against the fixed landmarks of the BraTS-Reg dataset. From left to right, mean absolute registration errors are shown for the initial dataset, affine, and the enhanced iRegNet, respectively. On each box, the red line is the median and the green triangle is the mean.

**Table 2.** Quantitative results of the proposed method and the baseline affine method on the BraTS-Reg challenge validation set. MAE denotes the average of median absolute error between the predicted coordinates and the ground truth coordinates, whereas Robustness represents the successful rate of measuring how many landmarks have improved MAE after the registration.

| Case | Initial MAE | Affine MAE | Affine Robustness | Enhanced iRegNet MAE | Enhanced iRegNet Robustness |
|---|---|---|---|---|---|
| BraTSReg_141 | 13.50 | 3.64 | 1.00 | 2.18 | 1.00 |
| BraTSReg_142 | 14.00 | 5.98 | 0.88 | 7.18 | 0.75 |
| BraTSReg_143 | 16.00 | 8.85 | 0.88 | 4.89 | 1.00 |
| BraTSReg_144 | 15.00 | 9.44 | 0.88 | 5.64 | 1.00 |
| BraTSReg_145 | 17.00 | 5.36 | 1.00 | 4.71 | 1.00 |
| BraTSReg_146 | 17.00 | 7.13 | 1.00 | 2.62 | 1.00 |
| BraTSReg_147 | 1.50 | 2.50 | 0.00 | 2.53 | 0.50 |
| BraTSReg_148 | 3.50 | 3.06 | 0.30 | 2.61 | 0.75 |
| BraTSReg_149 | 9.00 | 2.18 | 1.00 | 1.38 | 1.00 |
| BraTSReg_150 | 4.00 | 4.00 | 0.11 | 2.20 | 0.74 |
| BraTSReg_151 | 3.00 | 2.00 | 0.45 | 1.47 | 0.75 |
| BraTSReg_152 | 5.00 | 2.00 | 0.95 | 1.61 | 0.95 |
| BraTSReg_153 | 2.00 | 2.00 | 0.33 | 1.68 | 0.75 |
| BraTSReg_154 | 2.00 | 2.10 | 0.15 | 1.83 | 0.55 |
| BraTSReg_155 | 2.00 | 2.63 | 0.21 | 2.10 | 0.53 |
| BraTSReg_156 | 7.00 | 3.30 | 1.00 | 1.62 | 1.00 |
| BraTSReg_157 | 10.00 | 6.52 | 0.90 | 4.58 | 1.00 |
| BraTSReg_158 | 4.50 | 3.75 | 0.40 | 1.65 | 1.00 |
| BraTSReg_159 | 6.00 | 8.00 | 0.36 | 3.58 | 1.00 |
| BraTSReg_160 | 4.00 | 2.50 | 0.70 | 2.47 | 0.60 |
| Mean | 7.80 | 4.35 | 0.62 | 2.93 | 0.84 |
| StdDev | 5.62 | 2.46 | 0.36 | 1.63 | 0.19 |
| Median | 5.50 | 3.47 | 0.79 | 2.33 | 0.97 |
| 25quantile | 3.38 | 2.42 | 0.33 | 1.67 | 0.75 |
| 75quantile | 13.63 | 6.12 | 0.96 | 3.83 | 1.00 |

## 4    Conclusion

In this paper, we proposed a patient-specific registration framework based on iRegNet, which aligns pre-operative and post-recurrence MRI T1ce sequences. The enhanced iRegNet framework uses deep unsupervised learning for deformable image registration driven by the regularization hyperparameter. The proposed method is evaluated on the BraTS-Reg challenge dataset brain MR images comprising 140, 20, and 40 divided into training, validation, and testing cohorts. The validation results show that our framework can provide respectable results and is more effective than the classical affine registration with the advantage of being a self-supervised learning approach. In addition, iRegNet provided a faster approach (1 sec for 3D brain MRI pair registration) compared with conventional approaches than can last for minutes and hours in some tasks. The results clearly validate the effectiveness of using unsupervised deep-learning techniques in image registration.

Further research work should be conducted to investigate the optimal cropping radius for MRI sequences to minimize the missing data as possible. Automating this procedure will contribute toward rendering iRegNet an end-to-end pipeline. Further analysis of augmenting a supervised loss using weakly-supervised annotations and comparison against other deep learning methods is left for future work.

**Acknowledgments.** The first author is supported by the German Academic Exchange Service (DAAD) [scholarship number 91705803]. The authors acknowledge the help of the BraTS-Reg challenge organizers for submitting the Singularity container.